\documentclass[pra,twocolumn,floatfix,
amsmath,amssymb,aps,superscriptaddress,nofootinbib]{revtex4-2}
\usepackage{dcolumn,amsmath}
\usepackage[T2A]{fontenc}
\usepackage[utf8]{inputenc}
\usepackage{graphicx}
\graphicspath{ {./Images/} }
\usepackage{bm}
\usepackage{xfrac}
\usepackage{indentfirst} 
\usepackage{physics}

\usepackage{xcolor}

\usepackage{hyperref}
\hypersetup{colorlinks=true, citecolor=blue, urlcolor=blue, linkcolor=blue}

\usepackage{color}

\begin{document}

\title{Theoretical study of transition energies and matrix elements of the cadmium atom}


\title{Theoretical study of transition matrix elements in cadmium for vacuum-ultraviolet generation in $^{229}$Th nuclear clock applications}

\author{Gleb Penyazkov}
\email{penyazkovg10@mails.tsinghua.edu.cn}
\affiliation{State Key Laboratory of Low Dimensional Quantum Physics, Department of Physics, Tsinghua University, Beijing 100084, China}
\affiliation{Frontier Science Center for Quantum Information, Beijing 100184, China}

\author{Yanmei Yu} 
\email{ymyu@iphy.ac.cn}
\affiliation{Beijing National Laboratory for Condensed Matter Physics, Institute of Physics, Chinese Academy of Sciences, Beijing, 100190, China}
\affiliation{University of Chinese Academy of Sciences, 100049 Beijing, China}

\author{Leonid V.\ Skripnikov}
\email{skripnikov\_lv@pnpi.nrcki.ru,\\ leonidos239@gmail.com}
\affiliation{Petersburg Nuclear Physics Institute named by B.P. Konstantinov of National Research Centre
``Kurchatov Institute'', Gatchina, Leningrad District 188300, Russia}
\affiliation{Saint Petersburg State University, 7/9 Universitetskaya nab., St. Petersburg, 199034 Russia}

\author{Shiqian Ding}
\email{dingshq@mail.tsinghua.edu.cn}
\affiliation{State Key Laboratory of Low Dimensional Quantum Physics, Department of Physics, Tsinghua University, Beijing 100084, China}
\affiliation{Frontier Science Center for Quantum Information, Beijing 100184, China}
\affiliation{Beijing Academy of Quantum Information Sciences, Beijing 100193, China}

\begin{abstract}    

The relativistic Fock-space coupled-cluster methods are applied to the cadmium atom. A large number of transition energies and matrix elements are calculated for the $5s^{2}\:$$ ^{1}S \to 5snp\: ^{1,3} P^{o}$, $5s6s\: ^{1}S \to 5snp\: ^{1,3} P^{o}$ and $5s5d\: ^{1}D \to 5snp\: ^{1,3} P^{o}$ transitions for a wide range of $p$ states accounting for relativistic and electron-correlation effects. The results obtained within two different approaches (Fock-space coupled cluster and configuration interaction) are compared with available experimental and theoretical data. Good agreement is found between the two methods for transitions involving low-lying excited $p$-states, whereas for high-lying states, the discrepancy becomes large. The calculated values are used to determine the third-order nonlinear susceptibility of cadmium vapor, with an agreement within 5\% between the different methods. The results of the present computations support the feasibility of generating vacuum ultraviolet light in Cd vapor via a four-wave mixing process for the direct excitation of $^{229}$Th nucleus. 

\end{abstract}

\maketitle

\section{Introduction}

Cadmium atom vapor~\cite{xiao2024,Kung1972,Mitev1978,Heinrich1984,Miyazaki:84} has been selected as one of the nonlinear media for the generation of vacuum ultraviolet (VUV) laser light through four-wave mixing (FWM) method. However, to design and conduct the corresponding experiments, and to accurately estimate the power of the generated light, various atomic properties, such as transition rates between the states involved in the mixing scheme, are required. While these rates have been measured for a wide range of transitions in other alkaline-earth and divalent-like atoms, no experimental or theoretical data are available for transition rates involving high-lying states of neutral Cd~\cite{Nist2023}. 

The multireference valence-universal relativistic Fock-space coupled-cluster method (FS-CC) accurately accounts for both electron-correlation and relativistic effects with high accuracy and is powerful for calculating different electronic characteristics of heavy atomic and molecular systems~\cite{Eliav2024}. Accurate ionization potentials and excitation energies have been calculated for various atoms~\cite{Eliav2015FSCC,Eliav2015superheavy,Eliav2017,Sato2018,Oleynichenko2020_sym,Lu2022,Kaygorodov2022,Eliav2024}(see also references therein). However, to our knowledge, Fock-space coupled cluster method has not been used as extensively for calculating transition matrix elements as it has for transition energies.

Experimental estimates of transition dipole moments in Cd are only available for only a few transitions \cite{Nist2023}. Previously, knowledge of transition rates relied primarily on several theoretical studies, determining dipole moment transition matrix elements in cadmium. In Refs.~\cite{Dzuba2019,Yamaguchi2019}, the configuration interaction method with many-body perturbation theory (CI+MBPT)~\cite{CI-MBPT} was used to calculate the transition matrix elements for transitions from the ground $5s^{2}\:$$ ^{1}S$ and excited $5s5p\: ^{3} P^{o}$ states to the lowest-lying spin-allowed states, up to the $5s7s\: ^{3} S$ state. In the work~\cite{Yamaguchi2019} these values were also computed within a CI framework, combined with the all-order linearized coupled-cluster method~\cite{Safronova2009}. Later, in Ref.~\cite{Zhou:2021}, the Dirac-Fock plus core polarization~\cite{DFCP} and relativistic configuration interaction methods were employed to obtain transition matrix elements for the same set of transitions. In the recent work~\cite{Zhang2024}, the set of calculated transition matrix elements has been expanded within the relativistic CI+MPBT framework to include low-lying transitions from the $5s5p\: ^{3} P^{o}_1$ and $5s5p\: ^{3} P^{o}_2$ states. All of the aforementioned calculations use similar methodology and yield consistent results.  However, there is still no theoretical or experimental data available for transition matrix elements involving high-lying states in general.

In the present work, we apply the Fock-space coupled cluster method to perform calculation of the transition dipole matrix elements for the $5s^{2}\: ^{1}S \to 5snp\: ^{1,3} P^{o}$, $5s6s\: ^{1}S \to 5snp\: ^{1,3} P^{o}$ and $5s5d\: ^{1}D \to 5snp\: ^{1,3} P^{o}$ transitions with $n$ up to 14 for transitions from the $5s^{2}\: ^{1}S$ and $5s6s\: ^{1}S$ states and up to 10 with transitions from the $5s5d\: ^{1}D$ state. The calculations are also carried out with the CI+MBPT method. Good agreement between the two methods is found for a set of relatively low-lying transitions with $n$ up to 10, whereas calculations for transitions involving higher-lying states give rather different results. The agreement of the calculated transition energies and matrix elements for the transitions involving low-lying states validates the correctness of the performed computation. The obtained results are expected to be of great value for improving the efficiency of fourth-wave generation in the two FWM schemes previously employed~\cite{xiao2024,Kung1972, Mitev1978,Heinrich1984,Miyazaki:84}.

\section{Theory and calculation details}
\label{sec2}

\subsection{Fock-space coupled cluster}

For performing the calculations of the electronic structure of Cd, we used the relativistic version of the Fock-space multireference coupled cluster theory (see, for instance,~\cite{Lindgren1987,shavitt2009many,Visscher2001,Eliav2024} and references therein). In the present method, the exact multi-electron wave functions are expressed in terms of a wave operator $\Omega$ acting on model vectors $\tilde{\psi}_i$:
\begin{equation}
    |\psi_i \rangle = \Omega |\tilde{\psi}_i \rangle,
\end{equation}
where the wave operator takes a form
\begin{equation}
    \Omega = \bigl\{exp(\hat{T})\bigr\}.
\end{equation}
Here, curly brackets denote normal ordering of the operator, and $\hat{T}$ is the so-called cluster operator. This operator takes the form:
\begin{equation}
   \hat{T} = \sum_{k \leq n_h} \sum_{l \leq n_p} \hat{T}^{(k,l)},
\end{equation}
with
\begin{equation}
    \hat{T}^{(k.l)} = \sum_{\alpha \beta} t^{(k,l)}_{\alpha \beta} a^{\dagger}_{\alpha} a_{\beta},
\end{equation}
where $t^{(k,l)}_{\alpha \beta}$ is the cluster amplitude associated with the excitation $a^{\dagger}_{\alpha} a_{\beta}$, $a^{\dagger}_{\alpha}$ and $a_{\beta}$ denote sequences of creation and annihilation operators. The quantities $n_h$ and $n_p$ represent the numbers of holes and particles, respectively, for the target $n_h h n_p p$ sector of Fock space with respect to the Fermi vacuum.
By finding the solution of the equations with respect to the cluster amplitudes, one obtains the corresponding eigenvalues (i.e., energies) and wave functions for the multi-electron problem.

Atomic bispinors were obtained for this ion within a Dirac-Hartree-Fock calculation of the Cd$^{2+}$ core under the Dirac-Coulomb Hamiltonian. The Fock-Space coupled cluster method with single and double cluster amplitudes (FS-CCSD), within the Dirac-Coulomb Hamiltonian,  was employed.
We adopted the uncontracted Dyall's all-electron quadruple-zeta AE4Z basis set~\cite{Dyall:98,Dyall:06,Dyall:12}. To properly describe higher-lying states, the AE4Z basis set was augmented with 17 $s$-type, 18 $p$-type, 6 $d$-type, and 2 $f$-type diffuse functions as well as 4 $d$-type, and 2 $f$-type core functions. The total basis size was $50s43p27d14f6g3h$. We correlated all electrons of Cd and all the virtual orbitals in the FSCC calculations. The $0h0p$ Fock space sector corresponds to the Cd$^{2+}$ ion. The wave functions of the interest belong to the $0h2p$ sector. 
To avoid the typical problem of intruder states, the denominator shifting technique~\cite{Zaitsevskii^2017b} was used. This technique replaces the energy denominators arising in the equation for the cluster amplitudes $t^{(l,k)}_{\alpha \beta}$ with the shifted ones in the following way:

\begin{equation}
    D_K \longrightarrow D_{K}^{'}(n) = D_K + S_K \Big( \frac{S_K}{D_K+S_K} \Big)^n.
\end{equation}

Here $D_K$ is the initial energy denominator, $\mathnormal{S_K}$ is a shift parameter (in the present case, we resorted to real shifts~\cite{Zaitsevskii^2017b}) and $\mathnormal{n}$ is a non-negative integer parameter. Note that $\mathnormal{S_K}$ should be chosen to be relatively small compared to large $\mathnormal{D_K}$ values, while still being substantial enough to ensure that the shifted denominators remain negative and sufficiently far from zero to avoid divergences.
A very large active space was considered in order to obtain the highly lying excited states, including the $n=(5-9)s$, $n=(5-8)d$, and $n=(5-16)p$ particle orbitals, accounting for 110 active bispinors.
The minimum energy-shift parameters that allowed for convergence within this active space were $-$0.1 and $-$0.25 for the $0h1p$ and $0h2p$ sectors, respectively.
Previously, the denominator shifting technique was successfully developed~\cite{Zaitsevskii:2018b,Oleynichenko2020,IH2023} and employed~\cite{Penyazkov2022} to avoid divergences for electronic structure calculations in $0h2p$ Fock space sector. For a more detailed and comprehensive description of the technique, one may refer to Ref.~\cite{Eliav2024}.

To refine the results of the computation, we also performed calculation within a framework of the Intermediate-Hamiltonian (IH) Fock-space coupled cluster method for incomplete main model spaces~\cite{IH2023}. This approach involves partitioning the model space into a main model space and an intermediate model space, such that the contributions of the intermediate space functions to the target states are minimal, thereby avoiding the intruder-state problem. For more details, see Ref.~\cite{IH2023}. In this calculation, the total model space consisted of 224 lowest-lying particle orbitals, while the main 
active
space was chosen to comprise determinants with configurations $n=(5-7)s$, $n=(5-7)d$, and $n=(5-14)p$ in sector $0h1p$ and configurations $5sn$, where $n=(5-7),(5-7)d,(5-14)p$, in sector $0h2p$.

The basis set correction for energy calculation was estimated as follows: correlation calculation for the target states was performed using the equation-of-motion scalar-relativistic coupled cluster (EOM-RCC) within the modified AE4Z basis set (comprising of $48s42p27d14f6g3h$ functions) and modified AE4Z basis set with inclusion of even more higher-harmonic's functions (comprising of $48s42p27d14f8g7h5i$ functions). In this calculation, 2 $s$-type and 1 $p$-type core functions were removed, compared to the FS-CC calculation, to achieve convergence using CFOUR~\cite{Matthews:CFOUR:20} program package. The additional set of $g$,$h$, and $i$ functions was generated using the compact basis set construction technique~\cite{Skripnikov:13a,Skripnikov:2020e}. In these calculations, the scalar-relativistic variant of the exact two-component (X2C-1e) technique~\cite{IliasX2C} was used.
The basis set correction for transition energies was then calculated as a difference between calculation in two above-mentioned basis sets.

For the correlation calculation of transition matrix elements, the framework of the theory of effective operators~\cite{Hurtubise:93,Oleynichenko2023} was employed. This method implies the direct calculation of transition matrix elements, truncating the infinite exponential sum for the wave operator acting on the model space wave functions. In our calculations, we used quadratic truncation level~\cite{Oleynichenko2023,Zaitsevskii:2023}. The calculation of the property integrals was performed using a code developed in Ref.~\cite{Skripnikov:16b}. It carries out the calculation of property integrals over one-electron orbitals obtained in the Dirac-Hartree-Fock (DHF) calculation. The relativistic electron-structure calculations were performed using the {\sc dirac}~\cite{DIRAC19,Saue:2020}, Exp-T~\cite{Oleynichenko_EXPT,EXPT_website}, and CFOUR~\cite{Matthews:CFOUR:20} program packages.

\subsection{Configuration interaction}
The CI+MBPT method is a hybrid approach combining CI that takes into account an interaction between valence electrons and a method accounting for core-valence correlations using an MBPT operator \cite{Dzuba-PRA-1996,Berengut-PRA-2006}. In this work, the CI+MBPT calculation is implemented by using the AMBiT code \cite{AMBiT}. The first step is to solve the Dirac-Hartree-Fock (DHF) solution for core and valence electrons in the $V^N$, $V^{N-1}$, or $V^{N-M}$ approximations, where $N$ is the number of electrons and $M$ is the number of valence electrons. In either choice of potential, the resulting one-electron Dirac-Fock operator is
\begin{equation}
h_{DF}=c \vec {\alpha} \cdot \vec{p}+(\beta+1)c^2-\frac{Z}{r}+V^{DF},
\end{equation}
where
$ \vec {\alpha}$
and $\beta$ are Dirac matrices, and $V^{DF}$ is either of $V^N$, $V^{N-1}$, or $V^{N-M}$. The Breit and QED interactions are written into the Dirac-Fock operator. The Breit includes both Gaunt and retardation terms in the frequency-independent limit by
\begin{equation}
B_{ij}=-\frac{1}{2r_{ij}}(\vec{\alpha}_i \cdot \vec{\alpha}_j+ (\vec{\alpha}_i \cdot \vec{r}_{ij}) (\vec{ \alpha}_j \cdot \vec{r}_{ij} )/r^2_{ij}  ).
\end{equation}
The remaining valence and virtual orbitals are constructed as a linear combination of B-spline basis functions. In the next stage of the AMBiT code, the CI-space of CSFs are formed by allowing an arbitrary number of electron and/or hole excitations from a set of `leading configuration' up to some maximum principal quantum number, $n$, and orbital angular momentum, $l$. The AMBiT code implements an approach, referred as `emu CI' \cite{Geddes-PRA-2018}, that can reduce the computational cost associated with CI further. In the emu CI approach, in addition to the `large-side' CSFs, there are `small-side' CSFs, formed by allowing electron and/or hole-excitations from a set of leading configurations up to some $n$ and $l$ that can be less that those for large-side CSFs. Finally, the MBPT operator and corrections to energies are implemented in the AMBiT code that can include the core-valence interaction perturbatively.

In the Cd AMBiT calculation, the DHF calculation is carried out with $V^{N-2}$ potentials. The valence basis set is chosen to be $20spdf$ that includes $s$-, $p$-, $d$-, and $f$-orbitals with $n<20$. The $5s^2$ and $5s5p$ configurations consists of the set of leading configuration. The single and double excitations from the leading configurations up to $20spdf$ comprise the large-side CSFs of the CI space. The small-side CSFs are constructed by allowing the single and double excitations from the leading configurations up to $6sp5d4f$ in addition to the single excitation from those leading configurations up to $20spdf$. In all AMBiT calculations, the single and double excitations from the $4d$ core are considered and the MBPT calculations are performed with all one-, two-, and three-body MBPT diagrams included based on the $30spdfg$ basis.

\section{Results and discussion}
The calculated transition energies for the considered states are presented in Table~\ref{table:1}.

\begin{table*}
\caption{Electronic transition energies for the considered states relative to the ground state (in cm$^{-1}$). Column labeled ``FS-CCSD (shifts)'' shows the results obtained using the FS-CC method with regular energy-denominator shifts, column labeled  "FS-CCSD (IH)" shows the results obtained using the Intermediate-Hamiltonian FS-CC method. The ''Basis set'' column represents a correction to the transition energy accounting for more extensive basis set. For details see Section~\ref{sec2}.
}
\begin{ruledtabular}
\begin{tabular}{lcccccc}
State  & FS-CCSD (shifts) & FS-CCSD (IH) & Basis correction & Final (FS-CCSD) & CI+MBPT & Experiment~\cite{Nist2023}  \\
\hline
$5s5p\: ^{3}P^{o}_0$ & 29704 & 29679 & +32 & 29711 & 30126 & 30114 \\
$5s5p\: ^{3}P^{o}_1$ & 30263 & 30235 & +32 & 30267 & 30739 & 30656 \\
$5s5p\: ^{3}P^{o}_2$ & 31472 & 31435 & +32 & 31467 & 31862 & 31827 \\
$5s5p\: ^{1}P^{o}_1$ & 43674 & 43639 & +21 & 43660 & 43383 & 43692 \\
$5s6s\: ^{1}S_0$ & 52756 & 52725 & +81 & 52806 & 53806 & 53310 \\
$5s6p\: ^{3}P^{o}_0$ & 57864 & 57807 & +83 & 57890 & 58771 & 58391 \\
$5s6p\: ^{3}P^{o}_1$ & 57939 & 57882 & +83 & 57965 & 58854 & 58462 \\
$5s6p\: ^{3}P^{o}_2$ & 58132 & 58070 & +83 & 58153 & 59027 & 58636 \\
$5s5d\: ^{1}D_2$ & 58717 & 58673 & +66 & 58739 & 59566 & 59219 \\
$5s6p\: ^{1}P^{o}_1$ & 59489 & 59445 & +76 & 59521 & 60499 & 59907 \\
$5s7p\: ^{3}P^{o}_0$ & 64485 & 64420 & +87 & 64507 & 65364 & 64996 \\
$5s7p\: ^{3}P^{o}_1$ & 64512 & 64447 & +87 & 64534 & 65394 & 65026 \\
$5s7p\: ^{3}P^{o}_2$ & 64584 & 64516 & +87 & 64603 & 65490 & 65094 \\
$5s7p\: ^{1}P^{o}_1$ & 65023 & 64966 & +85 & 65051 & 65854 & 65501 \\
$5s8p\: ^{3}P^{o}_0$ & 67319 & 67251 & +88 & 67339 & 68197 & 67830 \\
$5s8p\: ^{3}P^{o}_1$ & 67332 & 67264 & +88 & 67352 & 68212 & 67842 \\
$5s8p\: ^{3}P^{o}_2$ & 67367 & 67298 & +88 & 67386 & 68245 & 67875 \\
$5s8p\: ^{1}P^{o}_1$ & 67566 & 67502 & +87 & 67589 & 68415 & 68059 \\
$5s9p\: ^{3}P^{o}_0$ & 68812 & 68743 & +89 & 68832 & 69684 & 69314 \\
$5s9p\: ^{3}P^{o}_1$ & 68820 & 68750 & +89 & 68839 & 69692 & 69321 \\
$5s9p\: ^{3}P^{o}_2$ & 68840 & 68769 & +89 & 68858 & 69711 & 69340 \\
$5s9p\: ^{1}P^{o}_1$ & 68947 & 68880 & +88 & 68968 & 69801 & 69439 \\
$5s10p\: ^{3}P^{o}_0$ & 69712 & 69642 & +89 & 69731 & 70562 & 70191 \\
$5s10p\: ^{3}P^{o}_1$ & 69716 & 69647 & +89 & 69736 & 70567 & 70196 \\
$5s10p\: ^{3}P^{o}_2$ & 69729 & 69659 & +89 & 69748 & 70579 & 70208 \\
$5s10p\: ^{1}P^{o}_1$ & 69793 & 69725 & +89 & 69814 & 70633 & 70267 \\
\end{tabular}
\end{ruledtabular}
\label{table:1}
\end{table*}

The results of the transition matrix element calculation using the FS-CCSD and CI+MBPT methods, along with their comparison to the existing theoretical and experimental data, are presented in Tables~\ref{table:2} and~\ref{table:3}. In Table~\ref{table:2} only the `strong' transitions (transition matrix element value is larger than 0.1 a.u.) are presented.

\begin{table*}
\caption{Calculated values of reduced E1 matrix elements (in a.u.) for neutral Cd and comparison to available theoretical and experimental data (the experimental values are deduced from the measured oscillator strengths).}
\centering
\begin{tabular}{lcccccccc}
\hline
\hline
 Transition & FS-CCSD (shifts)  & FSCCSD (IH) & CI+MBPT  &   Ref.~\cite{Dzuba2019} & Ref.~\cite{Yamaguchi2019}$^a$  &  Ref.~\cite{Zhou:2021}$^b$ &  Ref.~\cite{Zhang2024} & Experiment~\cite{Nist2023}\\
\hline
singlet $\to$ singlet & & & & & & & & \\
$5s^2\: ^{1}S_0 \to 5s5p\: ^{1}P^{o}_1$ & 3.462 & 3.462 & 3.552 & 3.435 & 3.440 & 3.4292 & 3.479 & 3.01(75) \\
$5s^2\: ^{1}S_0 \to 5s6p\: ^{1}P^{o}_1$ & 0.721 & 0.725 & 0.697 &  & 0.689 & 0.655 & 0.670 &  \\
$5s^2\: ^{1}S_0 \to 5s7p\: ^{1}P^{o}_1$ & 0.328 & 0.334 & 0.319 &  &  &  &  &  \\
$5s^2\: ^{1}S_0 \to 5s8p\: ^{1}P^{o}_1$ & 0.159 & 0.203 & 0.186 &  &  &  &  &  \\
$5s^2\: ^{1}S_0 \to 5s9p\: ^{1}P^{o}_1$ & 0.417 & 0.140 & 0.126 &  &  &  &  &  \\
$5s^2\: ^{1}S_0 \to 5s10p\: ^{1}P^{o}_1$ & 0.366 & 0.104 & 0.094 &  &  &  &  &  \\
singlet $\to$ triplet & & & & & & & & \\
$5s^2\: ^{1}S_0 \to 5s5p\: ^{3}P^{o}_1$ & 0.153 & 0.152 & 0.169 & 0.158 & & & 0.167 & 0.14(4) \\
\hline
singlet $\to$ singlet & & & & & & & & \\
$5s6s\: ^{1}S_0 \to 5s5p\: ^{1}P^{o}_1$ & 4.145 & 4.152 & 3.945 &  & & & 3.923 &  \\
$5s6s\: ^{1}S_0 \to 5s6p\: ^{1}P^{o}_1$ & 7.877 & 7.887 & 7.879 &  & & &  &  \\
$5s6s\: ^{1}S_0 \to 5s7p\: ^{1}P^{o}_1$ & 1.157 & 1.176 & 1.152 &  & & &  &  \\
$5s6s\: ^{1}S_0 \to 5s8p\: ^{1}P^{o}_1$ & 0.512 & 0.481 & 0.437 &  & & &  &  \\
$5s6s\: ^{1}S_0 \to 5s9p\: ^{1}P^{o}_1$ & 0.423 & 0.269 & 0.234 &  & & &  &  \\
$5s6s\: ^{1}S_0 \to 5s10p\: ^{1}P^{o}_1$ & 0.340 & 0.174 & 0.149 &  & & &  &  \\
$5s6s\: ^{1}S_0 \to 5s11p\: ^{1}P^{o}_1$ & 0.160 & 0.124 & 0.106 &  & & &  &  \\
singlet $\to$ triplet & & & & & & & & \\
$5s6s\: ^{1}S_0 \to 5s6p\: ^{3}P^{o}_1$ & 0.815 & 0.799 & 0.673 &  & & &  & \\
\hline
singlet $\to$ singlet & & & & & & & & \\
$5s5d\: ^{1}D_2 \to 5s5p\: ^{1}P^{o}_1$ & 5.562 & 5.587 & 5.353 &  & & & 5.427 & 6.23(1.56) \\
$5s5d\: ^{1}D_2 \to 5s6p\: ^{1}P^{o}_1$ & 12.128 & 12.162 & 12.296 &  & & &  & \\
$5s5d\: ^{1}D_2 \to 5s7p\: ^{1}P^{o}_1$ & 0.558 & 0.563 & 0.609 &  & & &  & \\
$5s5d\: ^{1}D_2 \to 5s8p\: ^{1}P^{o}_1$ & 0.403 & 0.217 & 0.232 &  & & &  & \\
$5s5d\: ^{1}D_2 \to 5s9p\: ^{1}P^{o}_1$ & 0.016 & 0.123 & 0.129 &  & & &  & \\
singlet $\to$ triplet & & & & & & & & \\
$5s5d\: ^{1}D_2 \to 5s5p\: ^{3}P^{o}_1$ & 0.267 & 0.269 & 0.200 &  & & &  & \\
$5s5d\: ^{1}D_2 \to 5s6p\: ^{3}P^{o}_1$ & 1.777 & 1.761 & 1.335 &  & & &  & \\
$5s5d\: ^{1}D_2 \to 5s7p\: ^{3}P^{o}_1$ & 0.313 & 0.312 & 0.296 &  & & &  & \\
\hline
\hline
\end{tabular}
\begin{flushleft}
    $^a$The values obtained within CI+All method, where available, are presented. \\
    $^b$The values obtained using high-order one-body and two-body core-polarization potentials are presented. \\
\end{flushleft}
\label{table:2}
\end{table*}

As seen in Table~\ref{table:2}, the present calculations agree well with the available theoretical and experimental data for the low-lying transitions. For the $5s^2\: ^{1}S_0 \to 5s5p\: ^{3}P^{o}_1$ transition, both FS-CCSD approaches give values closer to the experimental one, having a relative discrepancy of $\sim$8$\%$ with the experimental value, while the current CI+MBPT and previous calculations give a $\sim$15$\%$. In the case of $5s^2\: ^{1}S_0 \to 5s5p\: ^{1}P^{o}_1$ transition, all theoretical results agree within less than 3$\%$, while having a discrepancy of $\sim15\%$ with the available experimental value. In contrast, the $5s^2\: ^{1}S_0 \to 5s6p\: ^{1}P^{o}_1$ transition values given by both FS-CCSD methods differ by $\sim10\%$ compared to different versions of the CI method results, which give values with a discrepancy of less than 5$\%$. The largest discrepancy for this transition is found between the IH FS-CCSD and DFCP~\cite{Zhou:2021} methods. For the $5s6s\: ^{1}S_0 \to 5s5p\: ^{1}P^{o}_1$ transition, the present CI calculation agrees with the result from Ref.~\cite{Zhang2024} within a negligible deviation of less than 1\%. The difference with the corresponding FS-CCSD calculations is about 5\%. The values obtained for the $5s5d\: ^{1}D_2 \to 5s5p\: ^{1}P^{o}_1$ transition closely match other available theoretical results (with a difference of less than 5\%), with the FS-CCSD calculations being closer to the isolated experimental value.

\begin{table}
\caption{Calculated values of reduced E1 matrix elements (in a.u.) for neutral Cd.} 
\centering
\begin{tabular}{lccc}
\hline
\hline
 Transition & FS-CCSD (shifts) & FS-CCSD (IH) & CI+MBPT\\
\hline
$5s^2\: ^{1}S_0 \to$ & & & \\
singlet & & & \\
$5s11p\: ^{1}P^{o}_1$ & 0.089 & 0.083 & 0.074\\
$5s12p\: ^{1}P^{o}_1$ & 0.107 & 0.129 & 0.060\\
$5s13p\: ^{1}P^{o}_1$ & 0.170 & 0.171 & 0.051\\
$5s14p\: ^{1}P^{o}_1$ & 0.031 & 0.070 & 0.052 \\
triplet & & & \\
$5s6p\: ^{3}P^{o}_1$ & 0.005 & 0.005 & 0.009\\
$5s7p\: ^{3}P^{o}_1$ & 0.003 & 0.002 & 0.008 \\
$5s8p\: ^{3}P^{o}_1$ & 0.001 & 0.001 & 0.005\\
$5s9p\: ^{3}P^{o}_1$ & 0.033 & 0.0008 & 0.003\\
$5s10p\: ^{3}P^{o}_1$ & 0.023 & 0.0007 & 0.003\\
$5s11p\: ^{3}P^{o}_1$ & 0.012 & 0.0008 & 0.003\\
$5s12p\: ^{3}P^{o}_1$ & 0.036 & 0.0008 & 0.006\\
$5s13p\: ^{3}P^{o}_1$ & 0.028 & 0.0003 & 0.007\\
$5s14p\: ^{3}P^{o}_1$ & 0.001 & 0.0003 & 0.022 \\
\hline
$5s6s\: ^{1}S_0 \to$ & & & \\
singlet & & & \\
$5s12p\: ^{1}P^{o}_1$ & 0.026 & 0.108 & 0.080\\
$5s13p\: ^{1}P^{o}_1$ & 0.007 & 0.158 & 0.063\\
$5s14p\: ^{1}P^{o}_1$ & 0.059 & 0.216 & 0.060\\
triplet & & & \\
$5s5p\: ^{3}P^{o}_1$ & 0.087 & 0.087 & 0.091\\
$5s7p\: ^{3}P^{o}_1$ & 0.029 & 0.028 & 0.039\\
$5s8p\: ^{3}P^{o}_1$ & 0.006 & 0.005 & 0.005\\
$5s9p\: ^{3}P^{o}_1$ & 0.013 & 0.004 & 0.002\\
$5s10p\: ^{3}P^{o}_1$ & 0.010 & 0.005 & 0.003\\
$5s11p\: ^{3}P^{o}_1$ & 0.010 & 0.005 & 0.002\\
$5s12p\: ^{3}P^{o}_1$ & 0.029 & 0.008 & 0.002\\
$5s13p\: ^{3}P^{o}_1$ & 0.029 & 0.009 & 0.015\\
$5s14p\: ^{3}P^{o}_1$ & 0.008 & 0.004 & 0.033\\
\hline
$5s5d\: ^{1}D_2 \to$ & & & \\
singlet & & & \\
$5s10p\: ^{1}P^{o}_1$ & 0.008 & 0.072 & 0.085\\
triplet & & & \\
$5s8p\: ^{3}P^{o}_1$ & 0.165 & 0.090 & 0.097 \\
$5s9p\: ^{3}P^{o}_1$ & 0.042 & 0.003 & 0.055 \\
$5s10p\: ^{3}P^{o}_1$ & 0.026 & 0.006 & 0.038\\
\hline
\hline
\end{tabular}
\label{table:3}
\end{table}

If we analyse the results obtained for the transitions involving higher-lying states from Table~\ref{table:3}, we can see that, in general, two different coupled cluster approaches agree within $\sim15\%$ for transitions from all 3 considered states ($5s^{2}\: ^{1}S$, $5s6s\: ^{1}S$ and $5s5d\: ^{1}D$) to the states up to $5s8p\: ^{3}P^{o}_1$. In the case of transitions from the $5s^2\: ^{1}S_0$ state, the correspondence between the values for singlet $\to$ singlet transitions given by the FS-CC method with energy-denominator shifts (denoted as FS-CC (shifts)) and CI method is less than 15\% up to $5s8p\: ^{1}P^{o}_1$. For the IH FS-CC method, the agreement with a deviation of no more than 15\% with CI extends up to the $5s11p\: ^{1}P^{o}_1$ state. However, for the forbidden singlet $\to$ triplet transitions the situation is much less satisfactory: even for the $5s^2\: ^{1}S_0 \to 5s6p\: ^{3}P^{o}_1$ transition the values obtained by the coupled-cluster approaches differ by a factor of two compared to the CI method. Since the transition matrix elements for this transition are already quite small, we suggest that more thorough computations are needed for all higher singlet $\to$ triplet transitions. For the $5s6s\: ^{1}S_0 \to 5snp\: ^{1}P^{o}_1$ transitions, reasonable agreement is found up to $n=11$ when comparing IH FS-CCSD and CI+MBPT methods, and up to $n=8$ for FS-CCSD (shifts) and CI+MBPT methods. In contrast, for the $5s6s\: ^{1}S_0 \to 5snp\: ^{3}P^{o}_1$ transitions, the two methods yield results which differ by at least a factor of 2 for $n>8$. A 15\% agreement between the IH FS-CC and CI methods for singlet $\to$ singlet transitions is also observed for the transitions up to $5s5d\: ^{1}D_2 \to 5s10p\: ^{1}P^{o}_1$, although in the case of singlet $\to$ triplet transitions, the discrepancy of less than 10\% is achieved only for transitions up to the $5s5d\: ^{1}D_2 \to 5s8p\: ^{3}P^{o}_1$ transition, with a difference of about 25\% still present for the $5s5d\: ^{1}D_2 \to 5s5p\: ^{3}P^{o}_1$ transition.

Comparing the performance of the FS-CC (shifts) and IH FS-CC methods, it can be seen from Table~\ref{table:2} that a relatively good agreement between these implementations for singlet $\to$ singlet transitions is achieved up to the $5s^2/5s6s\: ^{1}S_0 \to 5s9p\: ^{1}P^{o}_1$ transition for those involving the $5s^2\: ^{1}S_0$ and $5s6s\: ^{1}S_0$ states and up to the $5s5d\: ^{2}D_2 \to 5s8p\: ^{1}P^{o}_1$ for those involving the $5s^2\: ^{1}S_0$ state. In these cases, the IH FS-CC method yields results that are generally closer to the CI+MBPT values, particularly for higher-lying transitions. For singlet $\to$ triplet transition, except for the transitions involving low-lying states up to the $5s8p\: ^{3}P^{o}_1$, all three methods give noticeably different results, with the values computed by IH FS-CC generally showing better agreement with those obtained from CI+MBPT. All considered methods agree with the experiment within the experimental uncertainty.

In the case of linearly polarized light, the third-order nonlinear susceptibility per atom, $\chi_{a}^{(3)}$, can be expressed as~\cite{AVSmith1987}

\begin{equation}
     \chi_a^{(3)}=\frac{1}{6 \epsilon_0\hbar^3}S(\omega_1+\omega_2)\chi_{12}\chi_{34},
\end{equation}
where
\begin{equation}
    \chi_{12}=\sum_l\left(\frac{\mu_{rl}\mu_{lg}}{\omega_{lg}-\omega_1}+\frac{\mu_{rl}\mu_{lg}}{\omega_{lg}-\omega_2}\right),
    \label{Chi12}
\end{equation}
\begin{equation}
    \chi_{34}=\sum_{m}\left(\frac{\mu_{rm}\mu_{mg}}{\omega_{mg}-\omega_4}+\frac{\mu_{rm}\mu_{mg}}{\omega_{mg}+\omega_3}\right),
    \label{Chi34}
\end{equation}
and $S(\omega_1 + \omega_2)$ describes the shape of the two-photon resonance.
Here, $\omega_{i}$ is the laser angular frequency, $\mu_{ij}$ are the $z$-components of the transition matrix elements coupling states $i$ and $j$. The summation runs over all intermediate states $l(m)$ (in the considered mixing schemes, these states are $n \: ^{1,3} P^{o}_1$), whose transition frequencies to the ground state $g$ are denoted as $\omega_{lg}(\omega_{mg})$.

Using the matrix elements calculated in the present work and the same laser frequencies ($\omega_1=\omega_2=375$~nm, $\omega_3=710$~nm), along with the same parameters for the shape factor $S(\omega_1 + \omega_2)$ for the two FWM schemes (using $5s6s\: ^{1}S_0$ or $5s5d\: ^{2}D_2$ as an intermediate state), considered in Ref.~\cite{xiao2024} for the direct excitation of the $^{229}$Th nuclear transition, the third-order nonlinear susceptibility $\chi_{a}^{(3)}$ of Cd vapor can be computed. The results obtained in the case of two-photon resonance for the pure $^{114}$Cd isotope are presented in Table~\ref{table:4}.

It can be seen that a notable agreement of 5\% is achieved for the nonlinear susceptibility $|\chi_{a}^{(3)}|$ and the corresponding factors $|\chi_{12}|$ and $|\chi_{34}|$ obtained using the IH FS-CC and CI+MBPT approaches for both considered FWM schemes. The values given by the FS-CC (shifts) method, however, while showing a 10\% agreement for the scheme utilizing the $5s5d\: ^{2}D_2$ intermediate state, exhibit a discrepancy of about 25\% compared to the IH FS-CC and the CI+MBPT methods for the $5s6s\: ^{1}S_0$ FWM scheme, thus, signifying the advantage of the IH FS-CC technique.

\section{Conclusions}

\begin{table}
\caption{Absolute values of the third-order nonlinear susceptibility of Cd vapor $|\chi_{a}^{(3)}|$ in units of $10^{-6}(ea_0)^4$cm$^{3}$ and the corresponding values of factors $|\chi_{12}|$ in units of $10^{-15}(ea_0)^2$Hz$^{-1}$ and $|\chi_{34}|$ in units of $10^{-16}(ea_0)^2$Hz$^{-1}$ for the pure $^{114}$Cd utilizing two FWM schemes with $5s6s\: ^{1}S_0$ and $5s5d\: ^{2}D_2$ as intermediate states.}
\centering
\begin{tabular}{lcccccc}
\hline
\hline
 & & FS-CCSD (shifts) && FS-CCSD (IH) && CI+MBPT\\
\hline
\multicolumn{7}{l}{ $5s6s\: ^{1}S_0$ scheme} \\
$|\chi_{12}|$ && 2.32 && 2.34 && 2.29 \\
$|\chi_{34}|$ && 4.67 && 6.34 && 6.29\\
$|\chi^{(3)}_{a}|$ && 1.10 && 1.50 && 1.46 \\
\hline
\multicolumn{7}{l}{ $5s5d\: ^{2}D_2$ scheme}\\
$|\chi_{12}|$ && 2.52 && 2.53 && 2.48 \\
$|\chi_{34}|$ && 6.91 && 6.57 && 6.27 \\
$|\chi^{(3)}_{a}|$ && 1.59 && 1.52 && 1.42 \\

\hline
\hline
\end{tabular}
\label{table:4}
\end{table}

We performed calculations of transition dipole matrix elements in neutral Cd for a large set of transitions using the Fock-space coupled cluster and configuration interaction with many-body perturbation theory (CI+MBPT) methods. In the case of Fock-space coupled cluster, two different implementations were used: the denominator shifting technique and the Intermediate Hamiltonian for incomplete model spaces. Previously, similar calculations for cadmium had been carried out only for a few low-lying transitions. Good agreement was found between the utilized methods and available theoretical and experimental results. A correspondence between the results obtained with the FS-CC and CI+MBPT approaches was achieved for a set of $5s^2\: ^{1}S_0 \to 5snp\: ^{1}P^{o}_1$ and $5s6s\: ^{1}S_0 \to 5snp\: ^{1}P^{o}_1$ transitions with $n$ up to 10. In the case of transitions to $5snp\: ^{3}P^{o}_1$ states, good consistency was observed for transitions with $n<9$. For the transitions involving the $5s5d\: ^{1}D_2$ state the results obtained with the FS-CC and CI methods are comparable for several low-lying $p$-states with $n<10$.

The values of the third-order nonlinear susceptibility obtained within the IH FS-CCSD and CI+MBPT methods agree within 5\% for both considered FWM schemes. Both methods account for electron-correlation effects in a different manner, thus, as a reference for the $|\chi^{(3)}_{a}|$, we suggest using the IH FS-CCSD results with an estimated uncertainty of 5\%.
These values are expected to be of significant interest for the preparation of neutral Cd spectroscopy experiment, as well as for the generation of laser light at 148.4 nm through four-wave mixing in cadmium vapor, aimed at direct excitation of the $^{229}$Th nucleus. The discrepancy between the FS-CC and CI results for high-lying transitions highlights the need for more advanced experimental and theoretical approaches to accurately determine considered properties.

\section*{Acknowledgments}
We thank Qi Xiao and Mikhail Reiter for discussions and comments.
Penyazkov and Ding are supported by the National Natural Science Foundation of China (No. 12341401 and No. 12274253) and the Shanghai Municipal Science and Technology Commission (No. 25LZ2600402), Yu is supported by the National Key Research and Development Program of China (2021YFA1402104), the Innovation Program 659
 for Quantum Science and Technology (Grant No. 660
 2021ZD0300901), and a Project funded by the Space Application System of China Manned Space Program. Skripnikov is supported by Russian Science Foundation under grant ~No.~24-12-00092, \url{https://rscf.ru/project/24-12-00092/}.
%

\end{document}